*Review*

# Synthesis of Modeling, Visualization, and Programming in GeoGebra as an Effective Approach for Teaching and Learning STEM Topics

**Rushan Ziatdinov** [1,*] **and James R. Valles, Jr.** [2]

[1] Department of Industrial Engineering, Keimyung University, Daegu 704-701, Republic of Korea
[2] Department of Mathematics, Prairie View A&M University, Prairie View, Texas 77446, USA; jrvalles@pvamu.edu
\* Correspondence: ziatdinov@kmu.ac.kr or ziatdinov.rushan@gmail.com

**Abstract:** GeoGebra is an interactive geometry, algebra, statistics, and calculus application designed for teaching and learning math, science, and engineering. Its dynamic interface allows its users to accurately and interactively visualize their work, models, and results. GeoGebra employs the synthesis of three key features: modeling, visualization, and programming (MVP). Many studies have shown the positive effects of GeoGebra on the efficiency and effectiveness of learning and teaching topics related to science, technology, engineering, and mathematics. In this study, we discuss how GeoGebra provides an environment for learning that is very interactive and collaborative between the learner and the instructor. We also show how integrating GeoGebra into the learning scheme can help improve the skills and knowledge of school and university students in numerous advanced mathematical courses, such as calculus, mathematical statistics, linear algebra, linear programming, computer-aided design, computer-aided geometric design, analytic and projective geometry, and graphical representation. Therefore, this study shows the effectiveness of GeoGebra and its MVP key features in science and engineering, particularly in topics related to mathematics. Each key feature of GeoGebra is thoroughly analyzed, and further analyses, along with how GeoGebra can be helpful in different topics, are discussed.

**Keywords:** GeoGebra; STEM; intelligent tutoring system; interactive learning environment; dynamic mathematics software; college mathematics; creative environment; modeling; programming; visualization

## 1. Introduction

According to Hohenwarter et al. [1], GeoGebra is an educational mathematics software program that conceptualizes and utilizes dynamic mathematics and is frequently used as a learning and teaching tool from middle school until the tertiary (postsecondary) level. GeoGebra was first presented in the school curriculum, but it was then expanded to include disciplines such as geometry, algebra, and calculus at the university level. GeoGebra is a software program designed for both teaching and learning, whose first and foremost goal is to make mathematical concepts clearer and easier for students to grasp. It is designed to enable proactive teaching and can, thus, be used to focus on problem-solving and assist with the development of mathematical experiments and concept introduction both in face-to-face and in remote class settings. With this program, learners can create sample problems of their own and then solve such problems using mathematical schemes and vital investigations. In this way, what would usually seem like taxing and daunting coursework and topics is given an appropriate avenue for proper, flexible, and supported exploration. This results in student learning that is based not on spoon-fed information, but rather on the learner's independence when they are making use of and honing their mathematical skills.

Besides this student-led exploration, dynamic worksheets are easily utilized in GeoGebra [1]. This makes it useful not only for middle- and high-school students, but also for college and university students, who can use it in more advanced mathematics classes, such as analytic and differential geometry and numerical methods. Overall, the dynamic preface of GeoGebra makes it very suitable and helpful for students at different educational levels. In some sense, GeoGebra can be regarded as a helpful tool for each and every learner in different mathematical areas. It allows for HTML exportation, which in turn allows for creative teaching kits for artistic visuals and mentoring aids, and this helps garner more class participation, which is widely attributed to dynamic worksheets [2].

In the current digital era, handheld devices such as mobile phones and tablets have become vital to the daily lives of individuals, students, and people in academia. Such transformation is widely due to the societal adaptability and mass acceptance of the continued digitalization of the current world. Mobile devices provide a reliable alternative for desktop computers in many ways, including in the conceptualization and understanding of mathematical discourse [3]. In comparison to computers, they are regarded as useful tools for cultivating proactive study places, hence proving their capability in bridging the gaps in learning mathematics at the university level, as opposed to relying on textbooks. They also help increase the focus and attention of learners through the visuals that they provide, along with the ease with which the students can send responses when they encounter a learning problem. One of such problems faced by younger generations is the increasingly short attention span. As an educational application, GeoGebra is highly recommended and can be downloaded from Google Play or Apple's App Store. It only requires the users to simply log in and allows them to publish and share their work with other users, thus making learning highly interactive and collaborative [3]. It also allows for the integration of technology into the academic curriculum, and it offers a collaboration of blended learning, both traditional and digital. Moreover, it helps university learners develop a deeper and more comprehensive understanding of mathematics, and its user-friendly interface allows learners to draw and simultaneously have algebraic functions that can all be entered directly with a keyboard, suiting the different learning needs of each learner.

GeoGebra is useful not only for students in middle and high school, but also for college students and educational instructors. In a study in 2019, Machromah et al. [4] tested the advantages of GeoGebra for university students studying and practicing calculus and found that it provides a substantial discourse. In another study, Haciomeroglu et al. [5] showed that GeoGebra helped the teachers familiarize themselves with the concepts of geometry, algebra, and calculus, which are commonly taught to university students. Particularly, it was found that the importance of representations, visualizations, and dynamic worksheets was imparted to the teachers [5]. Moreover, the courses that the teachers enrolled in helped them improve their skills to provide better and more up-to-date teaching aids and methods.

Overall, GeoGebra helps decrease the level of anxiety that many instructors might feel toward technology-integrated learning. To be able to effectively serve 21st-century learners, teachers need to modernize their pedagogical approaches, preferably with the current technological trends. In their research on GeoGebra's contributions to the professional development of mathematics instructors, Escuder and Furner [6] found that the teachers obtained experience particularly by navigating through the software and that this helped them improve their skills in order to productively pass on their knowledge to their students. Overall, a positive influence was linked to the mathematics teachers' perception of learning that is integrated with technology. Hence, GeoGebra helped increase their self-confidence in using technology in teaching.

Besides decreasing the level of anxiety and difficulty that teachers feel toward their digital literacy skills, GeoGebra provides a promising course of action in which instructors can practice and explore mathematics and subjects related to science, technology, engineering, and mathematics (STEM). It allows them to hone their digital literacy skills. Digital literacy refers to a person's capacity to find, assess, and clearly transmit information on a variety of digital platforms using typing and other media. According to the UNESCO, digital skills are defined as a set of abilities to access and manage information through digital devices, communication applications, and networks. Digital competence is one of the competences that educators need to instill in learners [7].

In general, GeoGebra can allow achieving two goals at once. That is, not only can it allow instructors to explore what technology has to offer and improve their own skills, but also it can allow them to hone their mathematical and STEM-oriented teaching skills. Ultimately, this can help instructors improve their self-efficacy and their relationship with their profession and with their students. Exploring GeoGebra allows teachers to improve their pedagogical knowledge and skills, thus providing them with an opportunity for an enhanced, more engaging, and more interactive learning atmosphere in the classroom.

In a study by Verhoef et al. [8], the authors showed the positive effects of GeoGebra on the professional development of instructors. This study focused on a four-year lesson study project conducted by teachers who were determined to elaborate and study derivatives with the help of GeoGebra. The study results showed how teachers coming from different Dutch schools explored and delved into calculus with the help of GeoGebra. They also showed how GeoGebra allowed these teachers to realize that conceptual embodiment and operational symbolism work hand in hand. Through a synthesis of GeoGebra and derivative consolidation, the teachers were able to zoom in on a graph of a derivative to infer and observe its local behavior [8]. In that study, the instructors tackled how they can acquire knowledge related to the importance of visualization and its significance to their students. Overall, the study findings revealed how teachers realized the advantages of GeoGebra in learning mathematics, particularly how GeoGebra helped them perceive the general learning process of their students.

The aim of this study is to assess the significance of GeoGebra to university-level learners. More importantly, our goal is to provide a synthesis of modeling, visualization, and programming (MVP), which are all achievable with GeoGebra's tools. In particular, in this study, we examine GeoGebra's success in teaching topics that are related to mathematics, such as science and engineering. Moreover, we provide a discussion of MVP along with a discussion of the problems that learners face pertaining to solving problems related to science and engineering.

## 2. Synthesis in Science and Engineering Education

In synergy [9], the whole is assumed to be greater than the sum of its parts. Synergy is important in synthesis formation, as it results in the fusion of many different yet parallel ideas of a certain topic or discourse. Similarly, according to Hampton and Parker [10], synergy is also a method for addressing concerns regarding information overload through ways that can aid in the development of scientific knowledge for decision-making. Overall, GeoGebra employs the synthesis or the integration of three key features: modeling, visualization, and programming (hereafter MVP). Integrating these features allows broadly tackling the importance of different mathematical fields. These features render the software successful and reliable.

In science, synthesis is what allows for the combination of research and case studies, which can turn into the fruition of new data and information [11]. According to Hampton and Parker [10], the importance of synthesis appears in how it aids in the flourishing of the scientific community, because it helps garner new partnerships and, to a greater extent, create new information. Synthesis is, thus, valuable in providing influence on policy and practice, but the interaction and collaboration of decision-makers in this context are widely needed [11]. Hence, effective synthesis has varying models, including governance models, situational arrangements, studies, and engagement operations.

Studies on mathematical learning and teaching are consistently performed to advance mathematics education. This research proves the significance and contribution of GeoGebra toward mathematics education. In creating a synthesis, mathematical beliefs have proven to be important [12]. As mentioned in [13], an in-depth synthesis has been used as a basis for studies on beliefs, impacts, and mathematical learning. For example, Malmivuori [13] examined the dynamic relationship between impact and cognition for learning operations present in mathematics education. A synthesis concerned with the affective domain, including causal attributions, self-efficacy, and self-confidence, as well as studies in this area, was found to relate to mathematical learning.

In line with the demands and capabilities of technology and student learning, GeoGebra is continuously packed and improved with various upgrades. For example, it is currently more easily accessible due to its cross-platform availability across different devices, such as laptops, desktop computers, and smartphones. Another improvement of GeoGebra is its automated reasoning tools on its most updated version. The fusion of the different mathematical and scientific-centric features provided by this software opens further possibilities for exploration and knowledge through its attributes, offering dynamic complexities.

Overall, GeoGebra can benefit the synthesis of science and engineering education. Academia involves prevalent fusion of science and engineering, which is particularly embodied in the field of STEM. A variety of topics and concepts covered by the STEM metafield are centered around and best learned through the synthesis of MVP. GeoGebra has all three of these features. Notably, the integration of engineering and mathematics is realized as a relevant force through the means of STEM-centered learning of mathematics. Therefore, GeoGebra comes into play in this matter as software in which independent learning takes place.

This independent learning was further demonstrated and deconstructed by Suweken [14]. In that study, observations and data showed that teachers in Indonesia lack the ability to successfully impart their mathematical knowledge to their students and that they still have much room for improvement. This encourages the application of mathematics and the integration of STEM subjects through the Standards for Mathematical Practice, which can bridge the gap in STEM-related activities [14]. In the study [14], GeoGebra mathematics applets (mathlets) were given to the students, who were then allowed to explore and test them and then showcase their results. These GeoGebra mathlets allowed the students to realistically apply, practice, and enhance their skills in mathematics and STEM education. They also helped them hone their skills and kindle their interest in the field, reflecting in turn on their teachers.

### 3. Modeling in GeoGebra

To provide a brief yet substantial context, it is worth mentioning that modeling in terms of mathematics is a key feature embedded in the process and knowledge of mathematics. According to Dundar et al. [15], mathematical modeling involves the process of conversion, which is a constant and inevitable part of realistic scenarios that involve mathematics. Modeling is the process of converting incidents that actually happen and converting mathematics-based incidents into an actual and realistic scenario.

Overall, GeoGebra can be considered a very creative tool for mathematical modeling. Doer and Pratt [16] defined two different types of modeling based on each student's task: exploratory modeling and expressive modeling. In exploratory modeling, a default model, made by a professional, is given to each user, thus enabling the learner to demonstrate their skills in model construction. While the model is being constructed, the student can stumble upon a pathway that results in a better understanding of the connection present amidst the model world and reality. According to Doer and Pratt [16], the modeling element found in GeoGebra provides a brilliant way of showing the cyclic view to the learners.

In the field of mechanical engineering, most of the topics constituting student courses involve the relative movement of various parts that make up a machine. Most books discussing mechanism and machine theory (MMT) provide a comprehensive description of assessments and combinations of connections along with approaches that are geometric in nature, which have withstood the test of time [17,18]. However, as the concepts of movement are explained in such books by static figures and arrows, these concepts can be difficult for students to grasp, especially when considering the fact that each individual has a different learning style. Hence, to bridge this gap, modern multimedia methods, such as videos, simulations, and other visual aids, are used, thus creating an interactive and encouraging learning environment [19]. One fundamental in MMT is dynamic geometry environments (DGEs), which allow the establishment of geometric structures.

GeoGebra includes a majority of the construction tools present in 2D computer-aided design software (Figure 1), which can help build dynamic models similar to those of geometric establishments with value-changeable specific parameters [19].

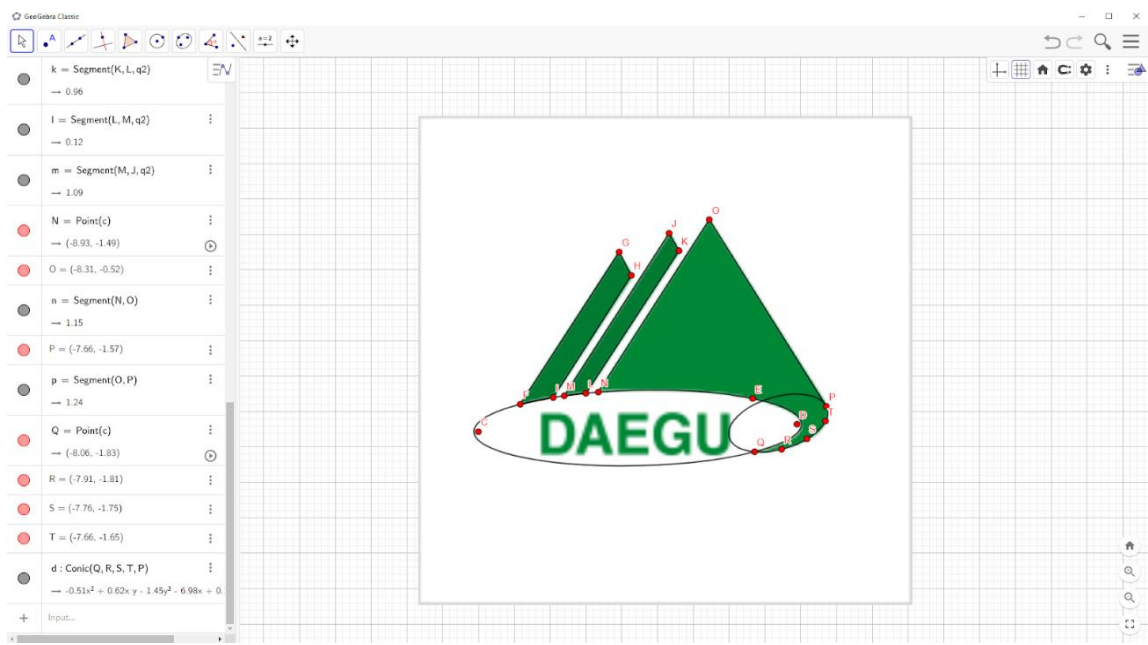

**Figure 1.** Using lines and conics to model the Daegu metropolitan city emblem. GeoGebra facilitates the calculation of the intersections of lines and conics.

Dynamic models are established in MMT, wherein parameters act as coordinates. Generally, numerous procedures in mechanism analysis are built upon geometry. In lessons based on theories following the concepts of geometry, related GeoGebra applets (available on the internet) are conveyed to supplement and fulfill students' understanding. These applets are interactive and can even be downloaded and modified to customize new tools [19]. GeoGebra improves the teachability of concepts in MMT, thereby allowing students to comply with requirements on time, with wide knowledge and application of MMT concepts.

Flehantov and Ovsiienko [20] performed a study to investigate the effectiveness of GeoGebra in gaining and cultivating knowledge on mathematical modeling. Overall, GeoGebra has its own spreadsheet feature, which allows its users to complete their requirements in portions in each study. The study results showed that GeoGebra provides a rapid formulation of intuitive spatial visualizations formed by students in terms of assessing the dynamic characteristics embedded in motion [20]. This was realized through the dynamic image formation of vector motion components. In this context, GeoGebra allowed for a great description and formulation of the mathematical schemas present in the software autonomously.

GeoGebra offers a constructivist teaching approach that allows students to independently build their own models while offering guidance [21]. It also makes the concept of oscillatory motion easier for learners to grasp. In this way, students can realize the desired outcomes of different models and simulations, hence enhancing their skills in science, technology, engineering, and mathematics (STEM), an aspect that is highly valuable in related careers. Modeling is highly required not only in scientific research, but also in studying physics [21]. GeoGebra allows the fusion of mathematics and physics through the advantages of information technology, hence providing students with a preview of the experience of creating models and simulations that pertain to physical phenomena [22,23]. With the students engaging in the establishment of models, their computational thinking skill is enhanced, which is a vital factor in STEM-related professions.

Generally, GeoGebra is continuously upgraded by its vast programming team. According to Mussoi et al. [24] and Aktümen and Bulut [25], GeoGebra allows model improvements, realistic phenomenon simulations, and realistic problem constructions. It also allows students to construct independently and to work in groups and apply their own knowledge in the field of mathematics.

GeoGebra provides engineering students with access to modeling parametric curves, such as trochoidal, epitrochoid, and hypocycloid curves [26]. According to Escuder and Furner [6] and Navetta [27],

Baravelle spirals and their relationship with infinite geometric series and complex numbers have been encapsulated. Moreover, Akkaya et al. [28] showed that GeoGebra is regarded as an avenue for students to study symmetry basic logic. GeoGebra also helps students learn statistics-related skills, including data management, analysis, and inference, as well as studying probability models [29]. It has also been found to be useful with the concepts of kinematics, quantum physics, thermal kinetics, and thermochemistry. According to Mussoi [24], GeoGebra aids in rationalizing rectilinear motion.

Many geographically specialized studies require the use of maps. One platform that is fundamentally useful in navigation is Google Maps. However, this platform still does not allow its users to designate points at certain locations or to show the distance between points [30]. GeoGebra allows the creation of a geometric topology on its interface, thus addressing this problem of Google Maps. It further allows the combination of geography, mathematics, and information technology visuals. This is achieved through a multistep construction, demonstrating the visual significance of the context of geography, and laying the foundation for future investigations [30].

## 4. Visualization in GeoGebra

### 4.1. Overview of Visualization

Since the beginning of time, visual imagery has been an efficient tool for transmitting both abstract and concrete ideas. Cave paintings, Egyptian hieroglyphs, Greek geometry, antique mosaics, and Leonardo da Vinci's groundbreaking methods of technical drawing for engineering and scientific purposes are all instances of this principle. Moreover, the introduction of computer graphics was pivotal in the development of scientific visualizations, hence influencing modern education at all levels [31].

Meditation can be used as an outlet for creative imagination, describing someone with a high level of confidence [32]. According to Spence [33], meditation is an inherently human task that employs cognition, far from computers. However, recently, visualization has been found to be related to data, computers, and humans' use of technology. According to Chen et al. [32], visualization is inherently connected to information quality. Through visualization, a data image is created, which helps relay information. This process also calls for converting one type of data to another. Hence, visualizing data allows many professionals, such as analysts and researchers, to form perceptions of data in a very straightforward manner, mainly because of the powerful skills connected to the visionary abilities of an individual, allowing them to observe and point out remarkable patterns quickly [34]. The quality of an image is undeniably linked to the skill of visualization. From a data-oriented point of view, visualization can be regarded as a means for accuracy and proficiency.

GeoGebra helps increase motivation and improve mathematical skills, self-awareness, and student learning involvement [35]. One of the main features of GeoGebra is its enhanced and supportive capability of visualization. In fact, visualization is considered a great and highly effective way for better learning. Learners can visualize algebraic problems using the dynamic graphics of GeoGebra (Figure 2). This promotes geometric thinking and provides algebraic and visual support for most learners [36]. It also allows learners to solve mathematical problems and to freely explore such problems globally and granularly for as long as they would like. According to Zimmerman [37], using representations and graphs as a problem-solving technique has proven helpful for obtaining correct and accurate answers.

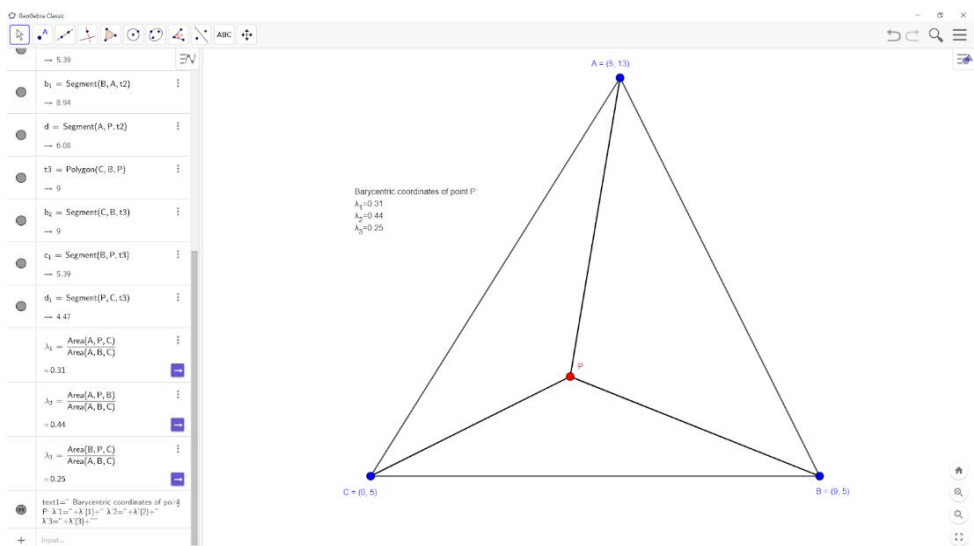

**Figure 2.** Computing the barycentric coordinates of point *P*. Rasterization, texture mapping, modeling surfaces, and performing point-in-triangle tests all employ barycentric coordinates in computer graphics.

Visualization offers an enhanced method of problem-solving. Visual representation, for example, enhances the process of learning as it allows for more interactions and manual application through formal definitions. Educational visualization includes creating a representation of a model or a process using a simulation so that it may be readily taught and explained. This is particularly useful when teaching topics that are difficult to visualize, such as atomic structure, temperature distribution, variation in the human body, the Gaussian curvature of a surface, zebra stripes on a composite surface, and a variety of other scientific and engineering-related issues. Educational visualization is, therefore, quickly enriching the subject of educational technology, which encompasses both theory and ethical practice in the educational process, across a variety of fields [38–44].

Overall, the autonomy that GeoGebra offers to every learner helps them develop an approach to mathematics and STEM-oriented subjects that is based on exploration and thinking instead of being solely focused on figuring out and coming up with a solution. This motivates the student to spend more time visually exploring a problem. GeoGebra also allows learners to use the exploratory approach and to cross-check their answers [45]. Through this key feature of GeoGebra, graphical representations of patterns and symbols allow students to visualize the task at hand, helping them gain insights through concepts rather than procedures [46]. The GeoGebra applets also allow accomplishing tasks that are mathematical in nature, because they aid operations connected to visualization, such as conjectures and exploration [47].

In general, the GeoGebra applets make the visualization process conducive. In particular, the dynamic feature of applets supports the exploration of a relative position in systems of linear equations. It enables many possible positions for two straight lines that can be drawn on a plane.

Marton et al. [48] explained the interrelatedness of visualization-associated processes and variation patterns. They first pointed out that there are hypotheses and conjectures that are linked to separation, a variation pattern intended to modify the core features of concepts. This is then followed by verification, which is fostered by variation pattern multiplication, an acting link between learners and mathematical elements [36].

*4.2. GeoGebra and Analytic Geometry*

Through its visualization features, GeoGebra aids in learning about geometry. Some of the elements required to obtain knowledge in geometry are the language of geometry, visualization skills, and effective instruction, which can each represent a barrier in learning the field of mathematics if not properly addressed [49]. With deficits in these elements, students encounter learning difficulties involving geometric topics. Idris [49] reiterated the importance of spatial visualization and its interrelatedness to successful knowledge in geometry, given that spatial visualization highly involves visualization itself. For example, a majority of students encounter difficulties with the visualization of three-dimensional objects through a two-dimensional point of view. These difficulties make it hard for them to truly appreciate the course in question. In their study, Saha et al. [50] investigated the effects of GeoGebra on learning geometry. They concluded that GeoGebra has positive effects on students learning geometry. Hence, GeoGebra is capable of providing different visualization levels to aid in developing the understanding of geometric ideas (Figure 3).

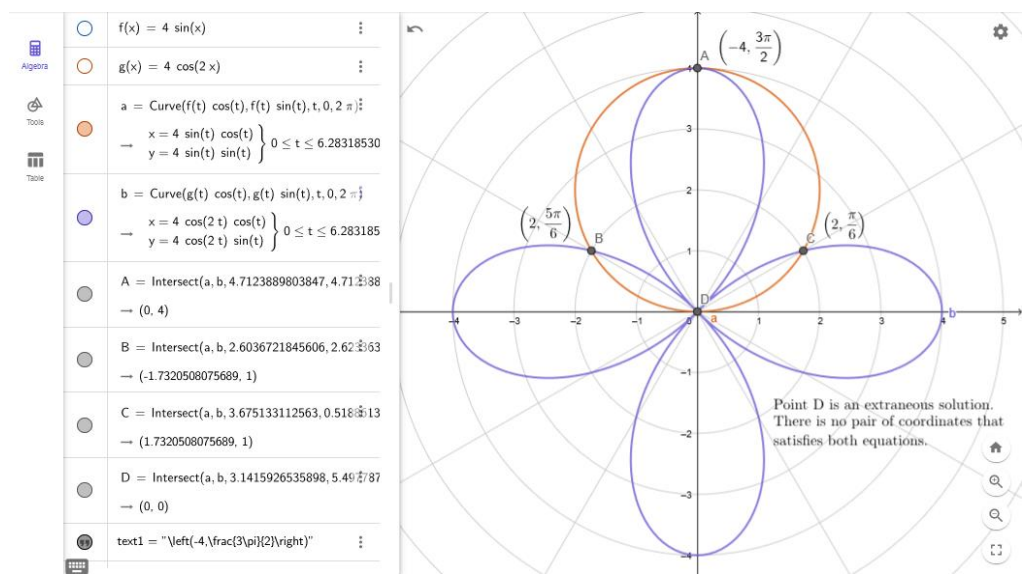

**Figure 3.** Finding the points of intersection of the graphs of two polar equations. The graphs can be easily obtained and the intersection points easily found, although it is necessary to algebraically verify the solutions found.

Symmetry is one of the vital concepts embedded in analytic geometry. In this context, GeoGebra is regarded as a suitable software program that can make learning the subject more approachable for students. GeoGebra allows for a constructivist method, which easily demonstrates point and line symmetries. Given how GeoGebra offers the function of actual graphics of much needed visuals for mathematical concepts and courses, such as that of symmetry, this makes GeoGebra suitable for sustaining the students' attention in learning otherwise daunting subjects, particularly because they are free to explore the concept on their own and, thus, provide realistic examples through GeoGebra. Subjects such as analytic geometry heavily rely on visuals and graphics to help learners actually demonstrate understanding and comprehensive knowledge that is practical and applicable, and this matches the functions that GeoGebra offers.

Trigonometry is a mathematical branch in which algebra, geometry, and graphical reasoning are all conjoined. Zengin et al. [51] performed a study to investigate the effects of GeoGebra on teaching and learning trigonometry. GeoGebra allows for the exploration of trigonometric functions and subtopics involving their graphs, which are considered essential elements in mathematics classes. In that study, the authors used a pre/posttest control group half-experimental pattern, and the results were found to exhibit some differences. These research results revealed that computer-assisted instruction, as an added guide for the constructivist teaching method, is more efficient than the constructivist teaching method alone [51].

Kepceoglu and Yavuz [52] explored the effectiveness of GeoGebra in learning and teaching trigonometry. Specifically, the aim of their study was to achieve this by teaching the periodicity of trigonometric functions. In this regard, they were determined to know and contrast the effectiveness of blended learning versus traditional learning on the students' understanding of the periodicity of trigonometric functions. The study results showed that the functions of GeoGebra, which allow for numerous representations of the periodicity of trigonometric functions, helped the experimental group reach correct and slightly correct answers [52]. Hence, it was concluded that the GeoGebra-assisted teaching approach for learning about the graphs of trigonometric functions in calculus is better than the traditional approach. Therefore, the study findings revealed that GeoGebra, which is used in an alternative teaching approach with the terms of periodicity of trigonometric functions that are typically taught algebraically rather than visually (using the traditional teaching approach), is a very effective teaching tool [52]. In turn, this helped clarify how numerous representations in concepts not limited to just trigonometry should be mandated in different courses and subjects.

GeoGebra can be used for studying space geometry in college. Pamungkas et al. [53] used a quasi-experimental design methodology to study the effectiveness of GeoGebra in the field of space geometry. They found that it has favorable effects on the knowledge of space geometry. Hence, GeoGebra is regarded as a viable tool in learning and teaching space geometry. It not only helps make the subject matter more theory-centered but also serves as a great aid by providing better learning methods as a teaching aid [53].

### 4.3. GeoGebra and Algebra

GeoGebra allows for graphic and visual representations of mathematical elements for topics that are rather tedious to teach through traditional methods, such as algebra. According to Dikovic [54], lesson plans that advocate for geometric representations allow teachers to provide guidance to their students while working and learning independently. In that study, Dikovic assessed and tested the applications of GeoGebra in teaching mathematics at the university level. The results confirmed that using the GeoGebra applets helped in the learning of differential calculus and significantly improved student learning and understanding.

GeoGebra is an effective tool that helps in the pursuit of algebraic knowledge. It represents an enhanced method that can bridge the gap in symbolic manipulation and computer algebra system (CAS) visualization, along with providing the flexibility of dynamic geometry software (DGS). GeoGebra has a special feature that is highly functional for algebra. Particularly, it is designed to have both of the functions that constitute DGS and CASs. In dual windows for algebra, students can work with points, lines, conic sections, automatic equations, and coordinates. Many studies have pointed out the effectiveness of using GeoGebra when working with concepts such as rational inequalities [55], linear equations and slopes [56], and exponential and logarithmic functions [57].

### 4.4. GeoGebra and Calculus

In general, the visualization concept required in both algebra and the geometrical aspects of calculus helps improve students' perceptions of derivatives. Sari et al. [58] studied how GeoGebra can aid in the exploration of concepts of derivatives through its dynamic visualization feature. They used the software to provide dynamic graphics of infinitesimal elements with respect to the concept of derivatives, specifically in terms of approaching the tangent line to a curve from a secant line. This helped establish a durable basis for students to be able to determine the derivative of a function in an algebraic way through its meaning. In this context, GeoGebra not only provided a visualization of a situation but also allowed reaching solutions for problems with different insights [58]. These findings prove that GeoGebra can allow students to turn their previous algebraic thinking into geometrical thinking while dealing with derivative problems.

Innovative learning models are generally useful in calculus. Many such models are made possible through GeoGebra [59], which provides this possibility through project-based learning models. According to Septian et al. [60], the project-based learning model aided by GeoGebra allows mentors to improve student learning opportunities by incorporating project tasks that can be accomplished with the software. Project work consists of complicated problem-based tasks that demand new information, based on real-life data, to be generated, accomplished independently by students or through groups, and concluded through GeoGebra project results. Moreover, the project-based learning model aided by GeoGebra helps students become more self-reliant in their studies, enhances their problem-solving skills, and maximizes the opportunities provided by technology [60]. For example, Septian [61] showed that students using a GeoGebra-assisted project-based learning model displayed improved mathematical representation capabilities.

Yimer and Feza [62] studied the integration of GeoGebra with the jigsaw cooperative learning strategy (JCLGS) for students learning calculus, focusing specifically on intermediate calculus for college students. Generally, many students exhibit a negative attitude toward calculus because of how challenging and intimidating it can be, with some even developing some type of phobia toward it [62]. These students, however, are not entirely to blame, as this sort of perception stems from the current learning and lecture setup, particularly the traditional method that relies on lectures. Overall, the experimental group of learners in that study [62] demonstrated positive results with JCLGS in learning calculus. They were able to actively visualize objects and materials in calculus that otherwise would have been theoretical and, therefore, difficult to put into realistic terms and understanding. This approach allowed them to achieve this because of the interactive features that GeoGebra offers, along with the flexibility that it offers in the representation of theoretical concepts that would otherwise be very difficult to emulate by the traditional method of lectures.

*4.5. GeoGebra and Linear Programming*

Linear programming is one of the core subjects of industrial engineering that is related to optimization. For a linear function subjected to multiple restrictions, linear programming is regarded as a mathematical modeling technique that maximizes or minimizes the function (Figure 4).

Hussen et al. [63] performed a study to determine the role that GeoGebra plays in making linear programming less complicated for students. Their study results showed that the teaching strategies used for linear programming lacked representation and that the traditional method is used with very little consideration for technological resources. Hence, GeoGebra allows linear programming materials to be more easily and accurately represented mathematically. It also helps foster an interaction between the students and the teacher.

In another study, Hobri et al. [64] examined how GeoGebra can assist a pixton and kelase web e-comic focused on linear programming that depends on two variables. They used the Thiagarajan model, also known as the 4D model, and subdivided it into four stages. They found that the learning media produced met the criteria for validity, efficacy, and practicality [64]. They also proposed a score that fits the very high category during the validation of the e-comic developed using pixton and kelase in relation to linear two-variable programs, which GeoGebra caters for. Overall, their study showed how GeoGebra is helpful in linear programming.

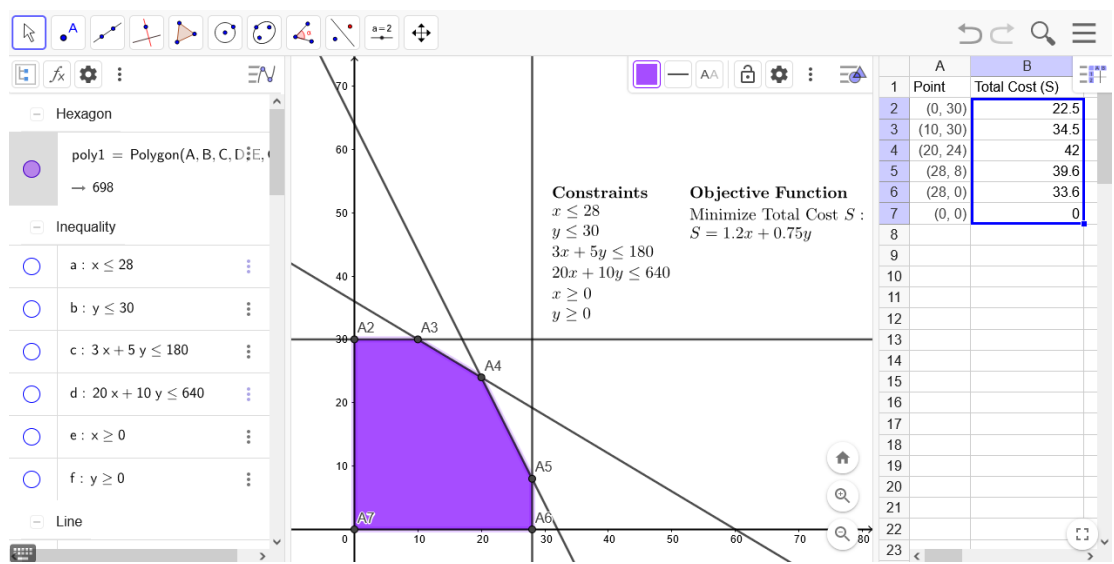

**Figure 4.** Solving a linear programming problem. Using the available linear graphing tools and the spreadsheet tool facilitates finding the feasible set, associated vertices, and optimal solution.

Visualization is an essential strategy for an individual to do better and be more efficient in mathematics. It allows the individual to have a clearer view of the task or problem at hand, which in turn allows them to have a better grasp of the problem instead of solely relying on a mental picture of it. In this context, visualization is a feature that GeoGebra caters for. This feature goes hand in hand with problem-based learning, which in turn helps improve the skills involved in linear programming.

Rahman and Johar [65] performed a study to assess how program-based learning (PBL) with GeoGebra increases the critical thinking skills of students. The results showed that GeoGebra is helpful in linear programming and in improving students' critical thinking skills in a given subject. This is because it provides the students with the autonomy required to critique and examine the graphic images that they freely draw within the settlement area. It also provides some level of autonomy with multiple possibilities of independently edited and modified graphics. Such features of GeoGebra allow individuals to think outside the box in their pursuit of learning and working with linear programming. Basically, GeoGebra helps in learning models that have mathematical problems as their basis. This helps students develop their critical thinking skills in linear programming.

## 5. Programming in GeoGebra

First and foremost, GeoGebra is an open-source software program, which allows it to run on desktop and laptop computers. A mobile version is also available. In the past, the GeoGebra applets used to run through a web browser Java plugin. However, since then, it has undergone many upgrades, eventually resulting in GeoGebraMobile. An HTML5 version called GeoGebraWeb is also available. However, unlike its predecessor, it no longer requires a Java plug-in. This allows all the resources of GeoGebra stemming from GeoGebraTube to be accessible on devices such as tablets and smartphones [66]. Both GeoGebraWeb and the desktop version of GeoGebra run on the same Java code base.

Recently, technology has turned into a necessity in every home, school, and corporation and elsewhere. Such increased use of technology in learning makes sense as the world continues toward ever-advancing modernity. The current generation of students prefer to have technology fused into their education. According to Abramovich [67], technology bridges the gap between comprehensive topics and dense topics, such as those found in mathematics. Hence, GeoGebra serves as a response to the needs of 21st-century learners in technology-integrated education.

In a study exploring the training of students in programming through GeoGebra, Velikova and Petkova [68] recruited as their participants 52 future mathematics and informatics teachers who were students at a university. They found that the training was successful when GeoGebra was used as a programming practice.

In general, instructors are encouraged to create quizzes and similar test types through GeoGebra by employing basic programming GeoGebra applets [69]. GeoGebra offers a dynamic learning environment that aids in the production of mathematical quizzes that are dynamic in nature. Overall, two types of scripts are used in GeoGebra: an internal GeoGebra script, wherein internal commands are embedded, and an external Java script [69]. It is also worth noting that GeoGebra has allowed the development of dynamic e-learning tools, such as student quizzes. In this context, a database that coincides with the applet's variable uses is created to store the data and information provided by the students [69]. Then, a line of communication is established between the students' data and the variables through a server and a database that the administrator(s) can access.

Generally, the programming features of GeoGebra can be integrated into the field of physics. This offers doable programming for physics simulations that can be used for teaching physics. Moreover, GeoGebra allows dynamic and flexible programming, which can be utilized in teaching physics. It also allows the production of simulations and animations, among other functions, even without prior knowledge of programming languages [70]. These simulations can run on different devices and do not require Java extensions. Additionally, GeoGebra allows the convenient exportation of animated GIFs for moving illustrations [70]. It also allows students and teachers to create physics simulations, which in turn allows them to gain a better understanding of the physics concepts that would otherwise require much effort to fully and comprehensively grasp.

With the continuous digitization of physics education, more tools are required. This was demonstrated by Solvang and Haglund [71], who examined how GeoGebra can help physics students in Sweden, particularly how it can aid their understanding of the subject. In their study, they emphasized that GeoGebra can facilitate an otherwise daunting topic by allowing simulations. For instance, it can help with problems dealing with polynomial functions. This is achievable because the software has a feature called sliders, in which the coefficients are entered. In summary, GeoGebra allows dynamic interactions between each and every representation, and this allows learners to uncover the links and basis for each factor in each representation [71].

## 6. Discussion and Conclusions

GeoGebra is a complex and dynamic mathematical software program that can aid in various mathematical calculations and tasks of varying degrees of complexity. It has been found to provide a significant number of advantages for different types of individuals, particularly for students, teachers, researchers, scientists, and mathematics professionals. It is easy to navigate through and to intuitively utilize and is, thus, convenient for its users. GeoGebra is very user-friendly and is easy to grasp, even for beginner users. Hence, it is regarded as a helpful tool for many different activities, such as studying, learning, and teaching. It is useful for not only elementary-school students, but also those in college and beyond, and it is also helpful and useful for their instructors. More importantly, it has the potential to bring about conciseness and interaction in learning about mathematics, science, and STEM-related topics and concepts through its dynamic features. It also brings to life the technology-integrated education that 21st-century learners need. In this context, it is important to highlight the work cited in [72], which provides different approaches, learning models, and strategies for building 21st-century skills in students during classroom activities with instructors and educators.

It is also important to highlight that, with the emergence of COVID-19 and its impact on education, there is a strong need among educators to learn how to integrate technology into their teaching. For instance, instructors should be aware of how to continue teaching when faced with circumstances that require virtual teaching with minimal preparation time. According to Beardsley et al. [73], many teachers believe that their abilities, confidence levels, and proficiency in using digital technology and creating digital content improved with the start of the COVID-19 lockdown. Digital content creation tools have become essential among Latin American university instructors [74]. However, it is worth mentioning that the recognition of the importance of technology as a tool in education is still insufficient. There is a balance that needs to be met with regard to students' use of technology and other aspects of their life [75]. Weinhandl et al. found that teachers who were capable of using technology in teaching were effective as schooling facilitators and that their students who regularly used technology in face-to-face settings effortlessly transitioned to home schooling with the same technology [76]. The researchers also found that using technology made learning about mathematics easier for the students, as technology was already an essential component before COVID-19 affected the learning environment. This finding supports the advantages of having teachers and instructors who are well trained and skilled in the use of technological tools for teaching and learning. It also highlights the versatility of GeoGebra in that it is well suited to meet the needs of mathematics students in many mathematical subjects.

Generally, GeoGebra offers several advantages in the fields of algebra, calculus, physics, and linear programming, among many other fields and subfields. It has continuously helped improve the performance, capabilities, and understanding of students due to its features, which allow learners to practice and visualize rather dense and complex topics. In this day and age, GeoGebra has a very high potential for this generation of learners as well as their learning and mental processes. Integrating GeoGebra into the current curriculum and teaching approaches has proven highly effective in various studies in different mathematical and STEM-oriented fields.

**Author Contributions:** Conceptualization, R.Z. and J.R.V.J.; methodology, R.Z. and J.R.V.J.; software, R.Z. and J.R.V.J.; writing—original draft preparation, R.Z. and J.R.V.J.; writing—review and editing, R.Z. and J.R.V.J.; visualization, R.Z. and J.R.V.J. All authors have read and agreed to the published version of the manuscript.

**Funding:** This research received no external funding.

**Institutional Review Board Statement:** Not applicable.

**Informed Consent Statement:** Not applicable.

**Data Availability Statement:** Not applicable.

**Acknowledgments:** We would like to express our gratitude to the reviewers for the time and effort spent to read the manuscript. We are grateful for all the helpful comments and recommendations that helped us improve the quality of the manuscript.

**Conflicts of Interest:** The authors declare no conflicts of interest.